\documentclass{emulateapj}

\usepackage{graphicx}
\usepackage{epsfig}
\usepackage{times}
\usepackage{natbib}
\usepackage{url}
\usepackage{color}
\usepackage{amssymb,amsmath}
\usepackage{hyperref}
\shorttitle{Fine structures of EUV waves}
\shortauthors{CHANDRA, CHEN, DEVI, JOSHI, SCHMIEDER, MOON, UDDIN}

\begin{document}

\title{Fine structures of an EUV wave event from Multi-viewpoint observations}

\author{Ramesh Chandra\altaffilmark{1}, P. F. Chen\altaffilmark{2,3}, Pooja Devi\altaffilmark{1},
Reetika Joshi\altaffilmark{1}, Brigitte Schmieder\altaffilmark{4}, Yong-Jae Moon\altaffilmark{5}, Wahab Uddin\altaffilmark{6}}

\affil{$^1$ Department of Physics, DSB Campus, Kumaun University, Nainital -- 263 001, India \email{rchandra.ntl@gmail.com}}
\affil{$^2$ School of Astronomy \& Space Science, Nanjing University, Nanjing 210023, China}
\affil{$^3$ Key Laboratary of Modern Astronomy and Astrophysics (Nanjing University), Ministry of Education, China}
\affil{$^4$ Observatoire de Paris, LESIA, UMR8109 (CNRS), F-92195 Meudon Principal Cedex, France}
\affil{$^5$ School of Space Research, Kyung Hee University, Yongin, Gyeonggi-Do, 446-701, Korea}
\affil{$^6$ Aryabhatta Research Institute of Observational Sciences (ARIES), Nainital 263 001, India}
\begin{abstract}
	In this study, we investigate an extreme ultraviolet (EUV) wave event on 2010 February 11, which 
occurred as a limb event from the Earth viewpoint and a disk event from the {\it STEREO}--B viewpoint. 
We use the data obtained by the Atmospheric Imaging Assembly (AIA) aboard the {\it Solar Dynamics Observatory} 
(SDO) in various EUV channels. The EUV wave event was launched by a partial prominence eruption. 
Similar to some EUV wave events in previous works, this EUV wave event contains a faster wave with 
a speed of $\sim$445$\pm$6 km s$^{-1}$, which we call coronal Moreton wave, and a slower wave with 
a speed of $\sim$298$\pm$5 km s$^{-1}$, which we call ``EIT wave". The coronal Moreton wave is identified as a fast-mode wave and the ``EIT wave" is identified as an apparent propagation due to successive field-line stretching.
We also observe a stationary front associated with the fast mode EUV wave. This stationary front is explained as mode conversion from the coronal Moreton wave to a slow-mode wave near a streamer.
\end{abstract}
\keywords{Sun: Prominence -- Sun: coronal mass ejection (CME) -- Sun: EUV waves}

\section{INTRODUCTION}

Since the discovery of extreme ultraviolet (EUV) waves, initially known as EIT waves named after the EIT telescope aboard the SOHO satellite, a number of studies have been done in order to clarify their nature. The readers are referred to the reviews \citep{Warmuth15, Chen16a} for details. Initially the studies were based on low-cadence (from several to more than 10 minutes) observations of SOHO and STEREO satellites \citep{Thompson99, Zhukov04, Zhukov09, Long17}. With these low-cadence observations, opposing viewpoints were proposed: Some groups proposed that EIT waves are fast-mode magnetohydrodynamic (MHD) waves, whereas the discovery of stationary wave fronts questioned this 
fast mode MHD wave interpretation \citep{Del99, Del2000, Zhukov09}. 
As a result, during the past decades several theoretical models were competing. These models include: fast mode wave model \citep{Vrs00,War04,Vrs08, War11},
slow-mode soliton model \citep{Wills07}, magnetoacoustic surface 
gravity waves \citep{Bal11}, magnetic rearrangement model \citep{Del99,Del2000}, current shell model \citep{Del08,Del14}, reconnection front model \citep{Attril07}, and a hybrid model where a faster component is attributed to fast-mode MHD waves and a slower component is due to successive stretching of closed magnetic loops \citep{Chen02, Chen05}.

After the launch of the SDO satellite in 2010, high temporal-resolution (up to 12 s) observation are available. Using these observations, \citet{Chen11} confirmed the simulations done by \citet{Chen02} and showed that there are two wave components in the time--distance plot of the EUV waves, i.e., a fast-mode MHD wave and a non-wave component. In order to avoid confusion, following \citet{Chen16a} we use ``coronal Moreton wave" for the fast-mode MHD wave component and ``EIT wave" for the slower non-wave component since we tend to believe that the fast EUV wave is the coronal counterpart of Moreton wave, whereas the EIT wave is produced by the successive stretching of magnetic field lines during flux rope eruption.

Due to their association with filament eruptions, coronal mass ejections \citep[CMEs, ][]{Chen11, Chandra16, Chandra18b}, type II radio bursts \citep{Nitta13, Chandra18a, Fulara19}, solar energetic particles (SEPs), and solar flares, the study of EUV waves becomes very interesting and crucial to understand the underlying physics behind these manifestations. One of the important consequences of EUV waves is to produce coronal loop oscillations \citep{Ballai05, Kumar13, Guo15, Fulara19} and filament/prominence oscillations \citep{Asai12, Shen14, Srivastava16}. These oscillatory phenomena are useful for coronal seismology in order to derive the coronal magnetic field, which is still challenging to measure. An important feature of the EUV waves is the occasional appearance of stationary brightening. We call this important because the discovery of these stationary fronts boosted the solar community to think about the EUV wave phenomenon in a different way. As a result, the idea of two types of EUV waves emerged and was verified after the availability of high-cadence SDO observations. This provides an opportunity for solar physicists to understand the EUV wave phenomenon in greater detail. Initially stationary fronts were associated with the EIT wave and were observed at magnetic separatrices. According to the magnetic fieldline stretching model \citep{Chen02, Chen05}, this is due to that the non-wave component of the EUV wave cannot travel from one magnetic domain to another, and so it stops at the location of magnetic separatrix between the two
magnetic domains. Magnetic separatrices are defined as the locations where the connectivity of magnetic field lines is discontinuous. \citet{Guo15} did a statistical study of twenty EUV waves and discussed different properties of stationary fronts associated with the non-wave component of EUV waves and reported that their speed varies from a few to 164 km s$^{-1}$. As found earlier the stationary fronts stop at magnetic separatrices, lasting for 500 to more then 2000 s. All these features can be explained by the magnetic fieldline stretching model.

However, in some recent observations, it was revealed that stationary fronts can also be created by fast-mode EUV waves \citep{Chandra16, Fulara19}. Using MHD simulations, \citet{Chen16b} reproduced this type of stationary fronts and proposed that they are different in nature from those found before \citep{Del99}. They are created by the fast-to-slow mode conversion of EUV waves at the places where the Alfv\'en speed is equal to the sound speed. Afterwards the resulting slow-mode wave stops at the location of magnetic separatrix and creates the observed stationary fronts. Very recently, more observations of this type of fast-to-slow mode conversion have been reported by \citet{Zong17} and \citet{Chandra18b}.

In this paper, we explore the kinematics and the dynamics of the EUV wave event observed on 2011 February 11. Together with this, stationary fronts associated with the fast EUV wave and the EIT wave are also studied. The paper is arranged as follows: We present the observations in Section \ref{observation}. The discussion is given in Section \ref{discussion}. Finally, we summarize our results in Section \ref{sum}.

\begin{figure*}
\centering
\includegraphics[width=0.7\textwidth]{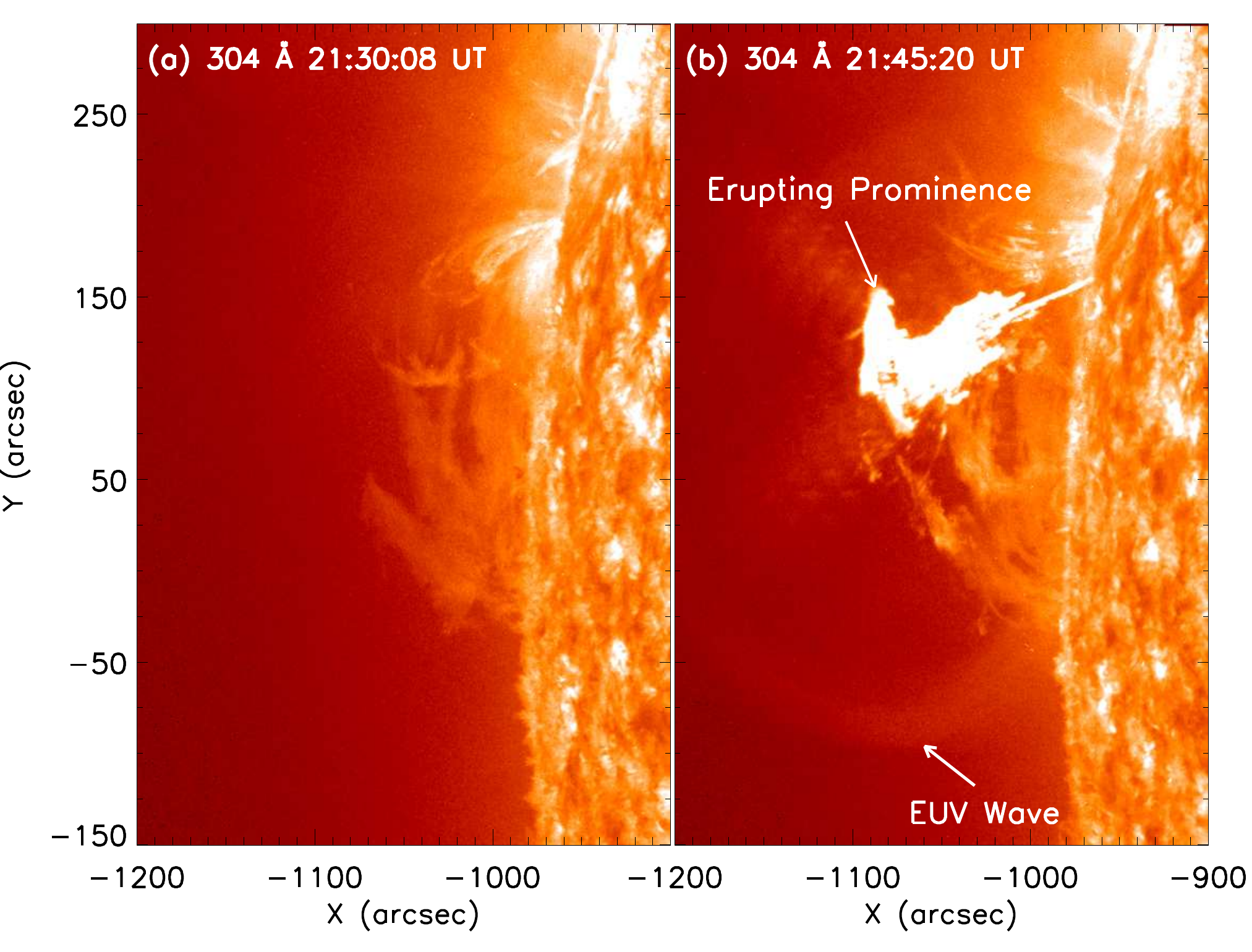}
\caption{Two snapshots of the event, one before prominence eruption (panel a)
and another during the eruption (panel b) in SDO/AIA 304 \AA. The leading
edge of the prominence eruption and the EUV wave are labelled
in panel (b).}
\label{fig1}
\end{figure*}
\section{OBSERVATIONS}\label{observation}

The present study contains the mutli–viewpoint and multi–wavelength observations of the EUV wave event on 2011 February 11. We use the data from the Atmospheric Imaging Assembly \citep[AIA,][]{Lemen12} onboard the Solar Dynamics Observatory \citep[SDO,][]{Pesnell12} satellite. Due to the full–disk high temporal and spatial resolution ultraviolet (UV) and extreme ultraviolet (EUV) observations of AIA, the dynamics of EUV wave events can be studied in greater detail. The pixel size and the temporal resolution of the full disk AIA data are $0\farcs 6$ and 12 s, respectively. We have aligned all the images using the routines available in Solar Software (SSW). For our analysis, we create difference images by subtracting the pre–event image from each EUV wave event image. The event is also observed by the Solar TErrestrial RElations Observatory--Behind \citep[STEREO--B,][]{Howard08}. From the perspective of STEREO--B the event is near the disc center. Therefore, we have an opportunity to explore this event with two different viewing angles from SDO and STEREO--B satellites. STEREO --B observes the full solar disk in four wavelengths, namely 304, 195, 284, and 171 \AA. The pixel size of the STEREO data is $1\farcs 6$. For our study, we use the 195 \AA\ waveband. The cadence of the STEREO 195 \AA\ data is 5 min.

The active region NOAA AR 11160, along with a prominence, is located at the east limb on 2011 February 11. The active region rotates to the west limb on 2011 February 25. During its appearance on the solar surface it produces several small-to-medium class solar flares. The flare on 2011 February 11 is registered as a GOES B8.0-class one and is associated with the partial eruption of the prominence. According to GOES observations the flare starts at $\sim$21:30 UT, peaks at $\sim$21:50 UT, and ends at $\sim$23:00 UT. A snapshot of the prominence eruption along with the associated flare is presented in Figure \ref{fig1}, which displays the evolution of the eruption in AIA 304 \AA\ waveband. Before eruption many threads of the prominence are visible above the east solar limb. However, only a part of the prominence erupts, which produces EUV waves together with a small flare. Therefore, this prominence eruption falls in the category of partial eruption as discussed in several past observations \citep{Chandra17, Zuca17}. The detailed review of this topic was presented in \citet{Parenti14}. The prominence eruption is associated with a CME with an angular width of 114$^\circ$. The CME becomes visible in the field-of-view of the LASCO C2 coronagraph around 22:12 UT. The linear speed as reported in the LASCO CME catalog is 469 km s$^{-1}$ and it decelerates with a rate of -6.3 m s$^{-2}$. This is a slow CME and no type II radio burst is registered.

\begin{figure*}[t]
\centering
\includegraphics[width=0.8\textwidth]{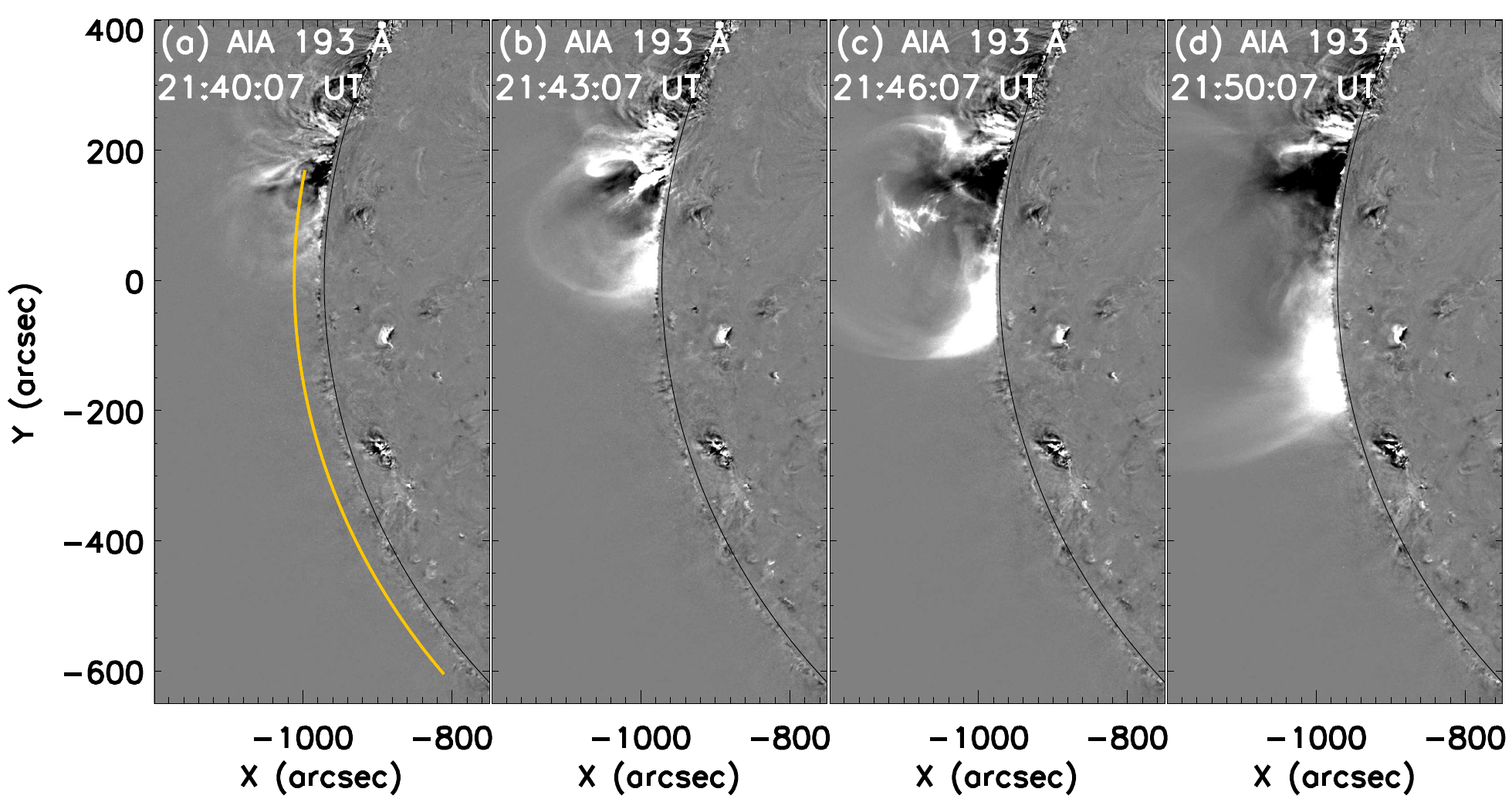}
\caption{Evolution of the EUV waves in SDO/AIA 193 \AA\ difference images.
The yellow curve in panel (a) represents the location of slice taken
for the computation of time-distance plot presented in Figure \ref{fig3}.
An animation related to this figure during 21:26 -- 22:10 UT is attached in 
the Electronic Supplementary Materials.}
\label{fig2}
\end{figure*}

\begin{figure}
\centering
\includegraphics[width=0.5\textwidth]{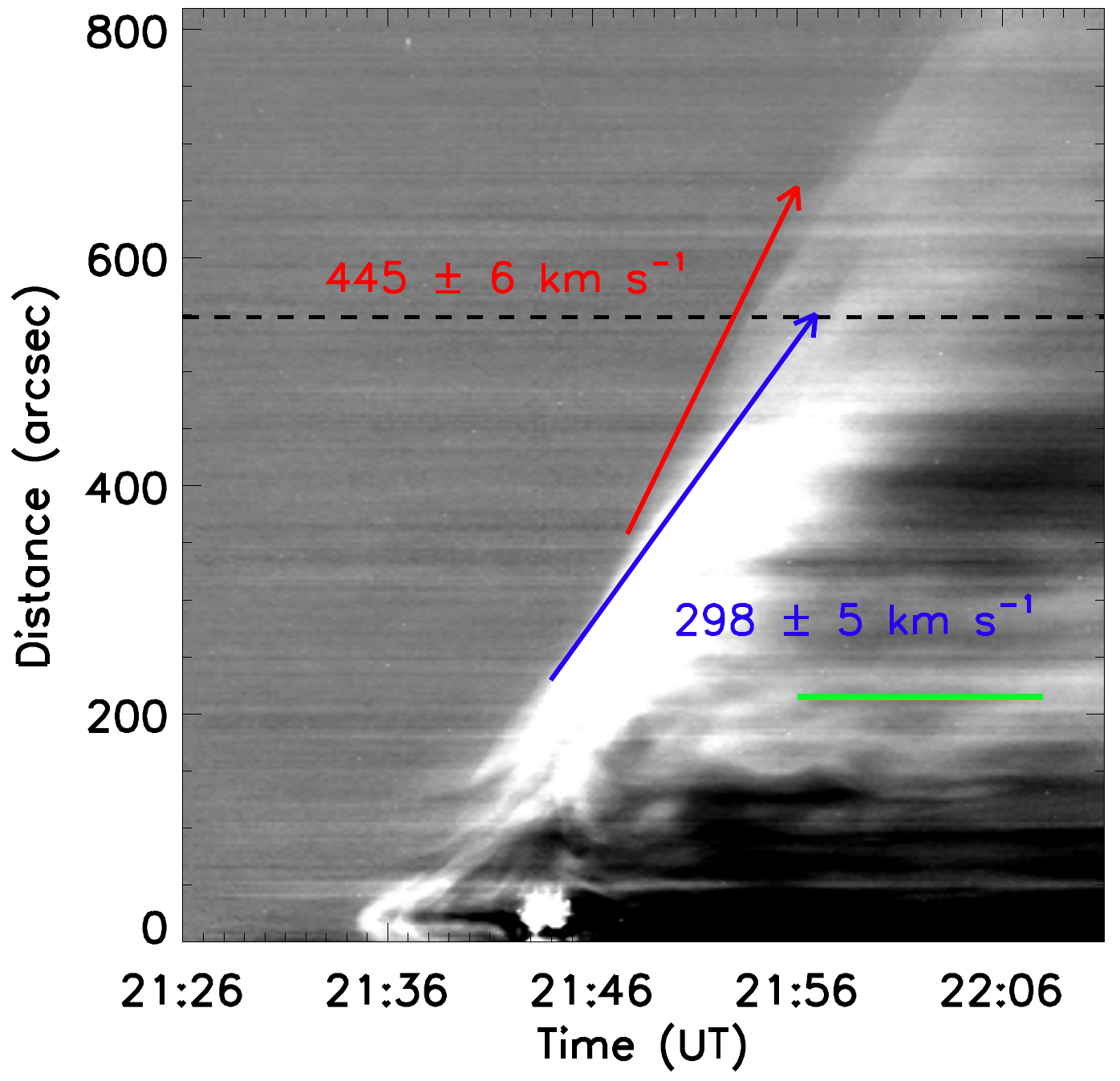}
\caption{Time-distance diagram along the yellow slice in Figure \ref{fig2}.
The coronal Moreton wave (red arrow), EIT wave (blue arrow) and a stationary
front (green line) are displayed. The intensity profile along the horizontal dashed 
line is used for plotting Figure~{\ref{fig4}}}.
\label{fig3}
\end{figure}

\begin{figure}
\centering
\includegraphics[width=0.5\textwidth]{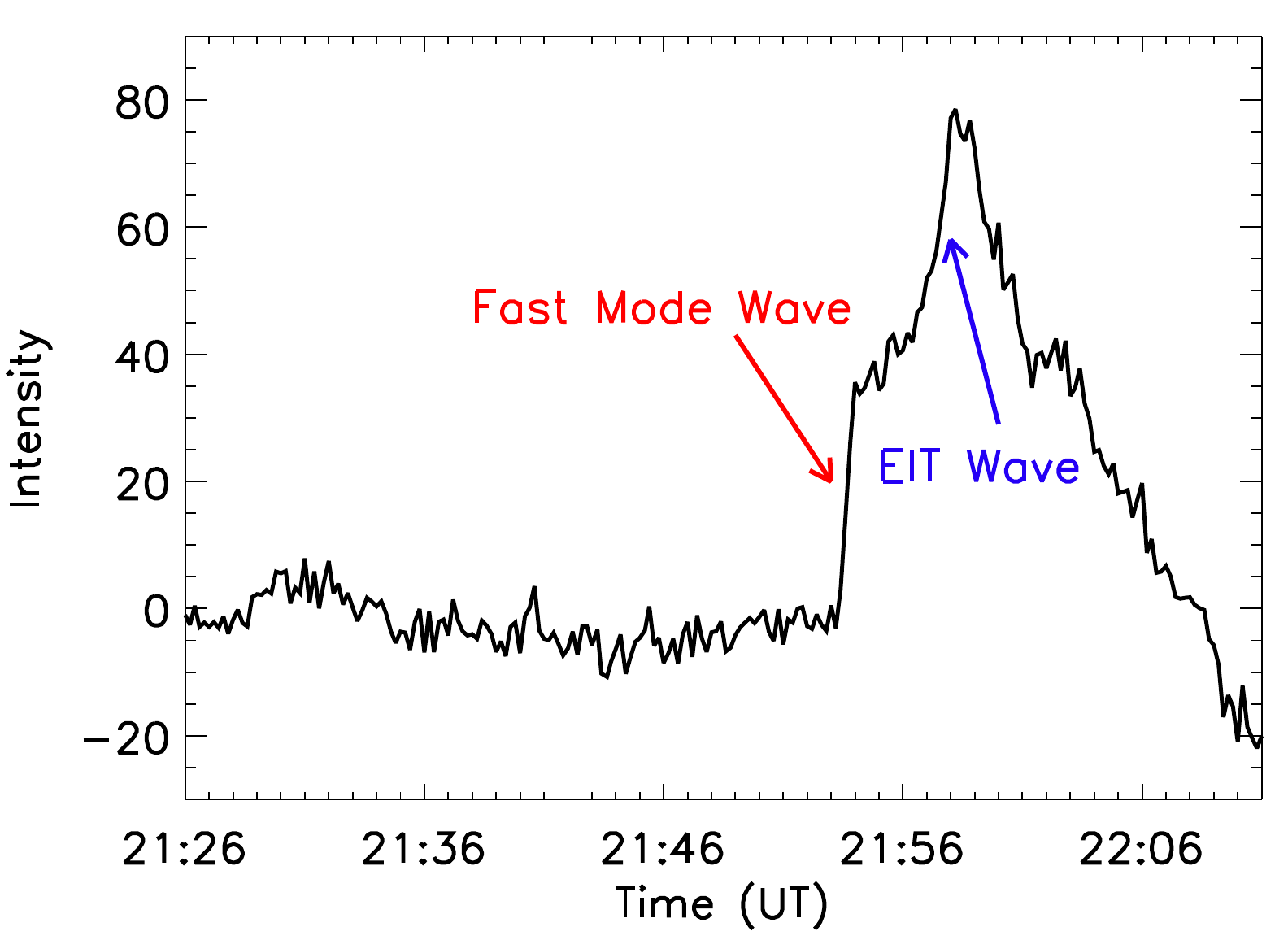}
\caption{Temporal evolution of the difference intensity along the horizontal dashed line marked 
in Figure~\ref{fig3}. The plot indicates the passage of two waves, i.e., a fast-mode MHD 
wave and a slower EIT wave.}
\label{fig4}
\end{figure}

The partial eruption of the prominence on 2011 February 11 produces EUV waves. The first appearance of an EUV wave
front is $\sim$21:37 UT in the SDO/AIA observations. Afterwards, the wave propagates along the south direction.
Figure \ref{fig2} shows four snapshots of the AIA 193 \AA\ difference images during the wave propagation.
The evolution of this wave event can be better seen in the attached movie.
It is seen that the wave appears with a dome shape in the AIA 193 \AA\ difference
images. For these difference images, we have subtracted the pre-event image at 21:26:07 UT.
To investigate the kinematics of the waves, we check the evolution of the 193 \AA\ difference intensity distribution along the yellow slice in Figure \ref{fig2}(a), which is parallel with the solar limb. The corresponding time-distance diagram of the 193 \AA\ difference intensity is plotted in Figure \ref{fig3}. It is seen that a bright front propagates away from the source active region as indicated by the blue arrow. The slope of the bright ridge characterizes the propagation velocity, which is 298 km s$^{-1}$
with an error of $\pm$5 km s$^{-1}$. The wave front is bright until $\sim$21:54 UT when the wave travels to a distance of 430\arcsec\ from the source region. After that, such a wave is still visible but with a much reduced intensity.
At the same time, another faint wave front can be identified, which is almost attached to the bright
front until 21:48 UT. Since then the faint front starts to separate further and further as indicated
by the red arrow. The faint wave front has a propagation velocity of 445 km s$^{-1}$
with an error of $\pm$6 km s$^{-1}$.

The estimation of the velocities of the two waves is based on the leading edges. The leading edges of both wave are defined by the local gradient maximum. In order to illustrate it, we extract a one-dimensional difference intensity profile at a distance of 548\arcsec\ from the time-distance data array of Figure \ref{fig3}, i.e., the difference intensity profile along the horizontal dashed line in Figure~\ref{fig3}. The profile is plotted in Figure~\ref{fig4}, which represents the evolution of the difference intensity at a distance of 548\arcsec\ from the eruption source region. In this figure, we clearly observe two sharp transitions, i.e., the first wave passes the location at 21:53:30 UT which is labelled ``fast-mode wave", and the second wave arrives at 21:57:30 UT which is labelled ``EIT wave". The first wave has the typical feature of a shock wave.

Besides the two EUV waves clearly identifiable in Figure \ref{fig3} is the presence of a stationary front at a distance of 200\arcsec. This stationary front is moderately bright and has a wide width. The location of this stationary front is shown by the green line in Figure \ref{fig3}.

\begin{figure*}[ht]
\centering
\includegraphics[width=0.9\textwidth]{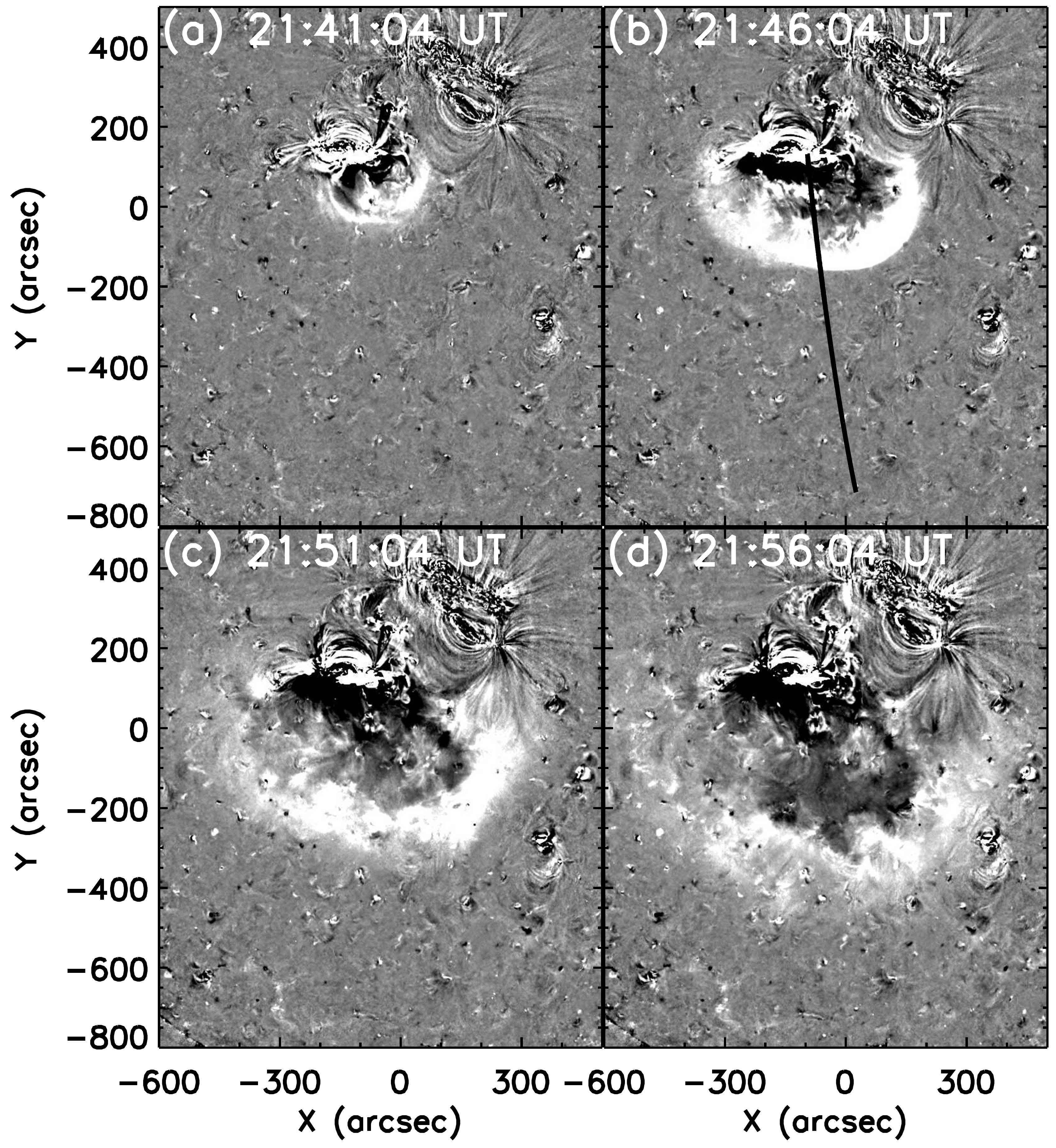}
\caption{Evolution of slowly moving EUV wave (EIT wave) observed by the STEREO B spacecraft in
195 \AA. Panel `b' indicates the location of the slice used for the time-distance
diagram in Figure \ref{fig6}. An associated animation of STEREO 195 \AA\ base difference 
images during 21:26 -- 22:06 UT is available.}
\label{fig5}
\end{figure*}

The event is also well observed by the STEREO--B telescope as a disk center event. The evolution of event observed by
STEREO 195 \AA\ is displayed in Figure \ref{fig5}, which displays the base-difference images at four times (see the attached movie).
 The base time is chosen to be 21:26:04 UT. It is seen that the eruption occurs near the southeast outskirt of the active region complex. As a result, wavelike perturbations are mainly visible in the southeastern quadrant. A bright front with less than half circle propagates out immediately followed by an expanding dimming. Perturbations are barely visible to the north of the source region.

\begin{figure}[t]
\centering
\includegraphics[width=0.5\textwidth]{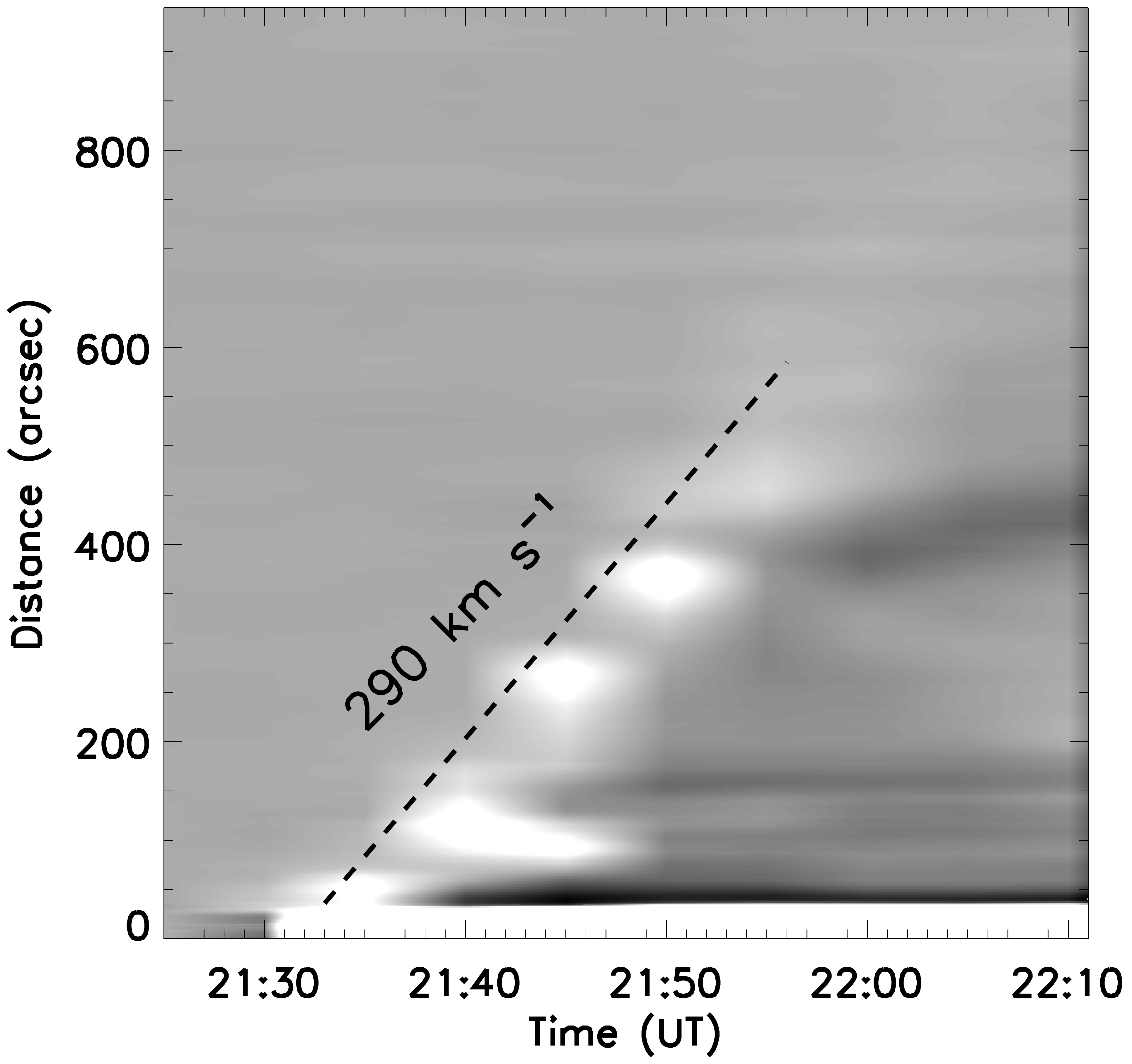}
\caption{Time-distance diagram along the slice displayed in Figure \ref{fig5}(b).}
\label{fig6}
\end{figure}

In order to compare the wave kinematic in STEREO 195 \AA\ and that in SDO 193 \AA, we determine the path of a slice in Figure \ref{fig5}(b), which corresponds to the east solar limb of the SDO view. The corresponding time-distance diagram of the STEREO 195 \AA\ difference intensity is shown in Figure \ref{fig6}.
The EUV wave observed from the top by STEREO has a
roughly constant velocity of $\sim$290 km s$^{-1}$.
It is noticed that the EUV wave observed by STEREO has a patchy feature, similar to the patchy ``EIT wave'' discovered by \citet{Guo15}. However, it should be pointed out that the ``patchy EUV wave'' in our Figure \ref{fig6} is an artifact, simply due to the low cadence of the STEREO, i.e., 5 min. With the data available, we can not tell whether patchy ``EIT wave'' appears in this event or not.

\begin{figure*}
\centering
\includegraphics[width=0.9\textwidth]{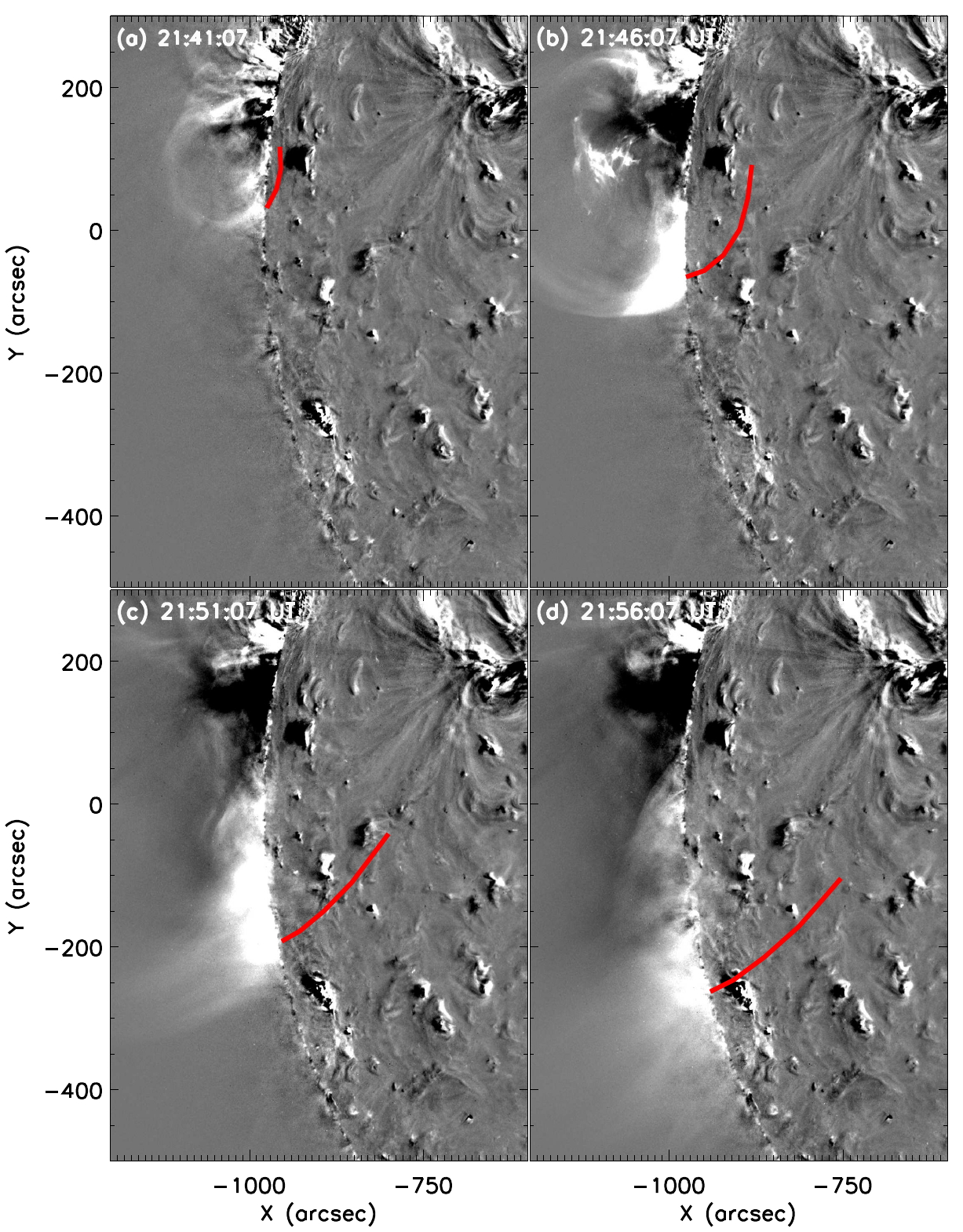}
\caption{Evolution of EUV wave in AIA 193 \AA\ difference images overlaid by the EIT
wave front observed by STEREO-B,
shown by red lines. The time of SDO image is written at the top left of each panel.
The EIT wave front are chosen at 21:41:04, 21:46:04, 21:51:04, and 21:56:04 UT for
panels (a), (b), (c), and (d), respectively, which are very close to the
AIA observations. The leading edges of the STEREO-B wavefront, which are determined 
by eyes, are overlaid on the SDO images.}
\label{fig7}
\end{figure*}

\begin{figure}
\centering
\includegraphics[width=0.5\textwidth]{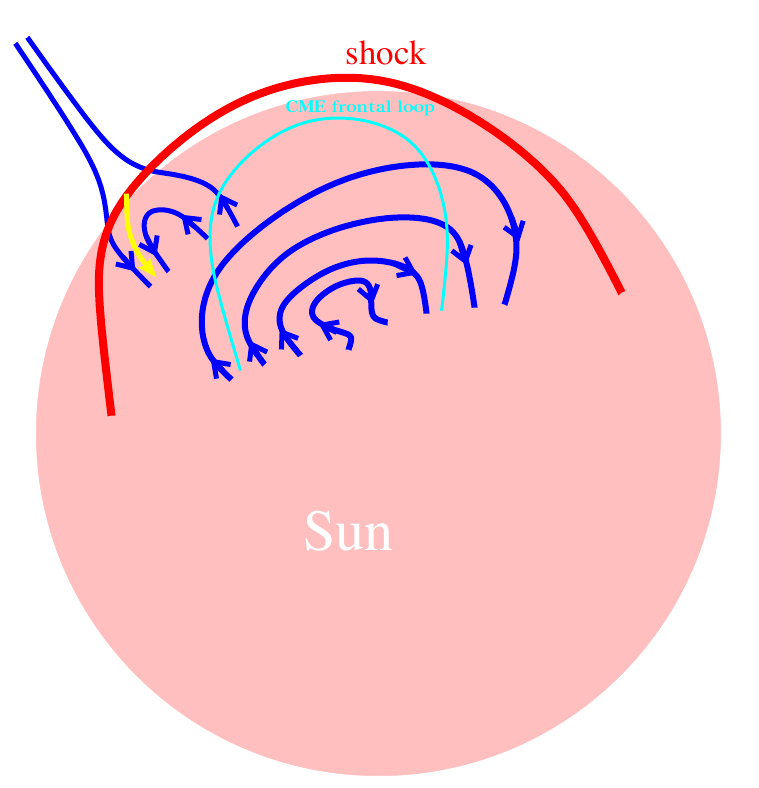}
\caption{Schematic sketch explaining the formation of a fast-mode shock (red), an ``EIT wave" (cyan) and a stationary wave front (yellow). As the flux rope erupts, a piston-driven fast-mode shock wave (red) is formed.
When it comes across the tip point of the streamer, part of the fast-mode wave is converted to a slow-mode wave.
At the same time, an ``EIT wave" (cyan), which also corresponds to the CME frontal loop, is formed due to successive stretching of the closed field lines above the erupting flux rope.}
\label{fig8}
\end{figure}

According to the time-distance diagram of the STEREO data, we detect only one bright EUV wave, rather than two EUV waves as in the SDO observations. In order to distinguish which wave in the SDO observations corresponds to the only EUV wave in the STEREO observations, we overlay the STEREO EUV wave fronts on the SDO 193 \AA\ difference images in Figure \ref{fig7}, where the background gray-scale images correspond to the SDO difference intensity maps, and the red lines correspond to the edges of the STEREO EUV wave fronts. This overplotting is done after transforming the STEREO coordinates to the AIA coordinates using the World Coordinate System available in the Solar SoftWare.
For the comparison, we use the SDO/AIA images at the times closest to each of the STEREO images.
We can see that the EUV wave front is still indistinguishable from the edge of the foremost SDO EUV wave front at 21:41:32 UT, but lags behind at later times. Therefore, the STEREO EUV wave should correspond to the slower one of the two EUV waves observed by SDO, i.e., the EUV wave detected by STEREO, or the ``EIT wave", rather than the coronal Moreton wave. This is reinforced by the almost identical velocities of the patchy STEREO EUV wave and the slower bright wave detected by SDO.
 The separation cannot be accounted for the projection effects as explained in the following imaginary 
experiment:
Suppose we are observing the AIA images in Figure \ref{fig7}
from the left side to mimic the STEREO viewing angle, the leading edge of the SDO wave front is also ahead of the STEREO wave front.
Moreover, if the faster faint wave and the slower bright wave in Figure \ref{fig3} are due to projection effects of a single dome-like wave, the intensity should vary smoothly from faint to bright. However, Figure \ref{fig3} indicates that it is a stepwise transition between the two waves.

\section{DISCUSSIONS}\label{discussion}

With the high cadence observations of the SDO/AIA data in various EUV wavebands, the nature of EUV waves, which were initially observed by the EIT telescope onboard the SOHO spacecraft, becomes clearer and clearer. Many groups have confirmed that the associated EUV waves in filament eruption events have two components, a faster one, which is a fast-mode MHD wave or shock wave, and a slower one, which is not a real wave \citep{Chen11, Asai12, White13, Chandra16}. The fast-mode EUV wave corresponds to the coronal counterpart of a Moreton wave observed in H$\alpha$ \citep{Chen16a}, which is expected in the model of \citet{Uchida68}. The slower one has been interpreted by different models. According to the hybrid model of \citet{Chen02, Chen05}, the faster one corresponds to the piston-driven wave or shock wave straddling above the erupting flux rope, and the slower one is an apparent structure that is generated by successive stretching of the closed magnetic field lines above the erupting flux rope, rather than a real wave. Therefore, two different names were called on for the two EUV waves, for example, ``coronal Moreton wave" for the faster one and ``EIT wave" for the slower one. According to this hybrid model, the coronal Moreton wave and the ``EIT wave" should exist in any filament eruption event. The main reason why the SOHO/EIT telescope detected one front is that the cadence of the EIT telescope was too low, e.g., $\sim$15 min. Another reason is that the other front is too faint to be observable. Recently, \citet{Zheng20} found that the activation of a filament without further eruption triggered a fast-mode MHD wave but not any slower wave behind. Such a result is also consistent with the magnetic fieldline stretching model since the magnetic field lines overlying the filament are not stretched out in the failed eruption case. In this paper, we analyzed the high-cadence SDO/AIA EUV data and verified the co-existence of two EUV waves, the coronal Moreton wave with a velocity of 445 km s$^{-1}$, and the ``EIT wave" with a velocity of 298 km s$^{-1}$.

However, one long-standing puzzle is that in many events only one EUV wave is detected even if the cadence is not so low. For example, when the fast-mode MHD wave is only about 400--500 km s$^{-1}$ so that the cadence of STEREO allows to catch the fast-mode MHD wave, it may still remain undetected. As we see in this paper, the limb observation by SDO indicates that a faint coronal Moreton wave is visible ahead of the bright ``EIT wave", traveling with a speed of 445 km s$^{-1}$. As implied by Figure \ref{fig7}, the coronal Moreton wave falls in the field of view of STEREO--B at several times. However, it is missing in the STEREO 195 \AA\ difference maps, where only the slower ``EIT wave" is identifiable. Therefore, this event provides an unprecedented opportunity to resolve the puzzle. The secret lies in the fact that the coronal Moreton wave is detectable in the SDO/AIA observations since the wave is observed from the side. The long line-of-sight along the fast-mode MHD wave front makes it easier to be detected. However, from the perspective of STEREO, the fast-mode MHD wave front is observed from the top, and the integrated EUV intensity is not strong enough to stand out from the background noise.

Statistical research indicated that the averaged Moreton wave speed, 664 km s$^{-1}$, is about 3 times faster than the averaged ``EIT wave" speed, which is about 205 km s$^{-1}$ \citep{Zhang11}. According to the magnetic fieldline stretching model \citep{Chen02}, if the magnetic field lines above the erupting flux rope are concentric semicircles, the coronal Moreton wave is indeed 3 times faster than the ``EIT wave'', well consistent with the observations. However, if the overlying magnetic field lines are more stretched in the vertical direction, the velocity ratio should be larger than 3. Similarly, if the overlying magnetic field lines are more stretched in the horizontal direction, the velocity ratio should be smaller than 3. In our event, the velocity ratio between the coronal Moreton wave and the ``EIT wave" is only about 1.5, implying that the magnetic field overlying the erupting prominence is less stretched in the vertical direction.

As mentioned in Section \ref{observation}, our Figure \ref{fig3} also indicates that besides the faster
coronal Moreton wave and the slower ``EIT wave", there is also a stationary EUV wave at a
distance of 200\arcsec\ from the source active region. In the SOHO era, most commonly
observed stationary EUV waves are explained in terms of sudden change of magnetic
connectivity \citep{Chen05}, i.e., the magnetic fieldline stretching stops at these sites.
The typical examples of the sudden change of magnetic connectivity are magnetic separatrices
or, more generally, magnetic quasi-separatrix layers \citep{Del99, Del2000, bala05}. However, very recently a few observations reported the formation of stationary fronts associated with fast-mode EUV waves \citep{Chandra16}. With MHD numerical simulations, \citet{Chen16b} proposed that such stationary EUV wave fronts are formed because part of the fast-mode MHD wave is converted to a slow-mode MHD wave when fast-mode MHD wave travels across a weak magnetic field area. According to \citet{Cally05}, such mode conversion happens when the local Alfv\'en speed is equal to the sound speed. The mode conversion of fast-mode MHD waves to slow-mode MHD waves was also reported in some other observations \citep{Zong17, Chandra18b}. The stationary front in our paper is generated when the coronal Moreton wave and the ``EIT wave'' are still too close to be distinguishable. This stationary front is not due to that the ``EIT wave'' stops near a magnetic separatrix since the ``EIT wave'' is still propagating. Therefore, it should be due to the mode conversion from the fast-mode coronal Moreton wave to a slow-mode MHD wave, which is then trapped inside a magnetic loop.

The weak magnetic field sites appear either around magnetic null points or well above a magnetic loop. Especially when the photospheric magnetic field is of multi-polarity, the coronal magnetic field decays rapidly with height, so frequently we can have a  site where the local \textbf{Alfv\'en} speed decreases to the sound speed.
To confirm our conjecture, we checked the SOHO/LASCO C2 coronagraph image and found that there exists a coronal streamer around the site. Since the magnetic field is very weak at the tip of streamers, such a place favors the condition that the local Alfv\'en speed is equal to the sound speed, as we found in \citet{Chandra18b}. Hence, a scenario in Figure \ref{fig8} is proposed here: As the prominence erupts, a piston-driven shock wave is generated straddling over the prominence, and a slower ``EIT wave" is produced by the successive magnetic field line stretching. When the widespreading fast-mode shock wave sweeps a coronal streamer behind the solar limb, a mode conversion happens, where part of the fast-mode wave is converted to a slow-mode wave. The slow-mode wave is trapped inside the magnetic loop, forming a stationary wave front seen in Figure \ref{fig3}.

\section{SUMMARY}\label{sum}

In this article, we present the observations of propagating EUV waves accompanied with a partial
prominence eruption on 2011 February 11. In a single event we found many strange features of EUV
waves, including fast-mode MHD wave, slower ``EIT wave" and
a stationary front associated with the fast-mode MHD wave. The main results are summarized as
follows: Two types of EUV waves with different velocities are clearly distinguished by the
AIA observations. To distinguish them, the faster one is called coronal Moreton wave and the
slower one is called ``EIT wave". Only the ``EIT wave" is nicely detected by the STEREO--B
satellite at the disk center from a different view angle. The locations of the ``EIT wave"
fronts are consistent very well between the SDO/AIA and STEREO--B observations.
A stationary EUV wave front is left behind the fast-mode coronal Moreton wave. We suggest that it is produced due to mode conversion when the fast-mode MHD wave travels through an area with weak magnetic field.

{\bf Acknowledgments}\\
We would like to thank the referee for the constructive comments and suggestions.
The authors thank the open data policy of the SDO team. P.D. is
supported by CSIR, New Delhi. P.F.C is finally supported by the the Chinese
foundations (NSFC 11961131002, 11533005, 275/2017/A and U1731241).
R.C. thanks P.F. Chen for his invitation to visit Nanjing University during October 2019.
B.S. thanks R.C. for his invitation to Nainital  in November 2019 and the discussions
on the movie showing the EUV wave. R.J. thanks to Department of Science and Technology 
(DST) for an INSPIRE fellowship.
\bibliographystyle{apj}
\bibliography{reference}

\begin{thebibliography}{}

\bibitem[\protect\citeauthoryear{{Asai} et~al.}{{Asai} et~al.}{2012}]{Asai12}
{Asai}, A., et~al. 2012, \apjl, 745, L18

\bibitem[\protect\citeauthoryear{{Attrill}}{{Attrill}}{2010}]{Attril07}
{Attrill}, G. D.~R. 2010, \apj, 718, 494

\bibitem[\protect\citeauthoryear{{Balasubramaniam} et~al.}{{Balasubramaniam}
  et~al.}{2005}]{bala05}
{Balasubramaniam}, K.~S., {Pevtsov}, A.~A., {Neidig}, D.~F., {Cliver}, E.~W.,
  {Thompson}, B.~J., {Young}, C.~A., {Martin}, S.~F.,  \& {Kiplinger}, A. 2005,
  \apj, 630, 1160

\bibitem[\protect\citeauthoryear{{Ballai}, {Erd{\'e}lyi}, \&
  {Pint{\'e}r}}{{Ballai} et~al.}{2005}]{Ballai05}
{Ballai}, I., {Erd{\'e}lyi}, R.,  \& {Pint{\'e}r}, B. 2005, \apjl, 633, L145

\bibitem[\protect\citeauthoryear{{Ballai}, {Forg{\'a}cs-Dajka}, \&
  {Douglas}}{{Ballai} et~al.}{2011}]{Bal11}
{Ballai}, I., {Forg{\'a}cs-Dajka}, E.,  \& {Douglas}, M. 2011, \aap, 527, A12

\bibitem[\protect\citeauthoryear{{Cally}}{{Cally}}{2005}]{Cally05}
{Cally}, P.~S. 2005, \mnras, 358, 353

\bibitem[\protect\citeauthoryear{{Chandra} et~al.}{{Chandra}
  et~al.}{2016}]{Chandra16}
{Chandra}, R., {Chen}, P.~F., {Fulara}, A., {Srivastava}, A.~K.,  \& {Uddin},
  W. 2016, \apj, 822, 106

\bibitem[\protect\citeauthoryear{{Chandra} et~al.}{{Chandra}
  et~al.}{2018a}]{Chandra18a}
{Chandra}, R., {Chen}, P.~F., {Fulara}, A., {Srivastava}, A.~K.,  \& {Uddin},
  W. 2018a, Advances in Space Research, 61, 705

\bibitem[\protect\citeauthoryear{{Chandra} et~al.}{{Chandra}
  et~al.}{2018b}]{Chandra18b}
{Chandra}, R., {Chen}, P.~F., {Joshi}, R., {Joshi}, B.,  \& {Schmieder}, B.
  2018b, \apj, 863, 101

\bibitem[\protect\citeauthoryear{{Chandra} et~al.}{{Chandra}
  et~al.}{2017}]{Chandra17}
{Chandra}, R., {Filippov}, B., {Joshi}, R.,  \& {Schmieder}, B. 2017, \solphys,
  292, 81

\bibitem[\protect\citeauthoryear{{Chen}}{{Chen}}{2016}]{Chen16a}
{Chen}, P.~F. 2016, Washington DC American Geophysical Union Geophysical
  Monograph Series, 216, 381

\bibitem[\protect\citeauthoryear{{Chen} et~al.}{{Chen} et~al.}{2016}]{Chen16b}
{Chen}, P.~F., {Fang}, C., {Chandra}, R.,  \& {Srivastava}, A.~K. 2016,
  \solphys, 291, 3195

\bibitem[\protect\citeauthoryear{{Chen}, {Fang}, \& {Shibata}}{{Chen}
  et~al.}{2005}]{Chen05}
{Chen}, P.~F., {Fang}, C.,  \& {Shibata}, K. 2005, \apj, 622, 1202

\bibitem[\protect\citeauthoryear{{Chen} et~al.}{{Chen} et~al.}{2002}]{Chen02}
{Chen}, P.~F., {Wu}, S.~T., {Shibata}, K.,  \& {Fang}, C. 2002, \apjl, 572, L99

\bibitem[\protect\citeauthoryear{{Chen} \& {Wu}}{{Chen} \& {Wu}}{2011}]{Chen11}
{Chen}, P.~F.,  \& {Wu}, Y. 2011, \apjl, 732, L20

\bibitem[\protect\citeauthoryear{{Delann{\'e}e}}{{Delann{\'e}e}}{2000}]{Del2000}
{Delann{\'e}e}, C. 2000, \apj, 545, 512

\bibitem[\protect\citeauthoryear{{Delann{\'e}e} et~al.}{{Delann{\'e}e}
  et~al.}{2014}]{Del14}
{Delann{\'e}e}, C., {Artzner}, G., {Schmieder}, B.,  \& {Parenti}, S. 2014,
  \solphys, 289, 2565

\bibitem[\protect\citeauthoryear{{Delann{\'e}e} \& {Aulanier}}{{Delann{\'e}e}
  \& {Aulanier}}{1999}]{Del99}
{Delann{\'e}e}, C.,  \& {Aulanier}, G. 1999, \solphys, 190, 107

\bibitem[\protect\citeauthoryear{{Delann{\'e}e} et~al.}{{Delann{\'e}e}
  et~al.}{2008}]{Del08}
{Delann{\'e}e}, C., {T{\"o}r{\"o}k}, T., {Aulanier}, G.,  \& {Hochedez}, J.-F.
  2008, \solphys, 247, 123

\bibitem[\protect\citeauthoryear{{Fulara} et~al.}{{Fulara}
  et~al.}{2019}]{Fulara19}
{Fulara}, A., {Chandra}, R., {Chen}, P.~F., {Zhelyazkov}, I., {Srivastava},
  A.~K.,  \& {Uddin}, W. 2019, \solphys, 294, 56

\bibitem[\protect\citeauthoryear{{Guo}, {Ding}, \& {Chen}}{{Guo}
  et~al.}{2015}]{Guo15}
{Guo}, Y., {Ding}, M.~D.,  \& {Chen}, P.~F. 2015, \apjs, 219, 36

\bibitem[\protect\citeauthoryear{{Howard} et~al.}{{Howard}
  et~al.}{2008}]{Howard08}
{Howard}, R.~A., et~al. 2008, \ssr, 136, 67

\bibitem[\protect\citeauthoryear{{Kumar} et~al.}{{Kumar}
  et~al.}{2013}]{Kumar13}
{Kumar}, P., {Cho}, K.-S., {Chen}, P.~F., {Bong}, S.-C.,  \& {Park}, S.-H.
  2013, \solphys, 282, 523

\bibitem[\protect\citeauthoryear{{Kwon} et~al.}{{Kwon} et~al.}{2013}]{Kwon13}
{Kwon}, R.-Y., {Ofman}, L., {Olmedo}, O., {Kramar}, M., {Davila}, J.~M.,
  {Thompson}, B.~J.,  \& {Cho}, K.-S. 2013, \apj, 766, 55

\bibitem[\protect\citeauthoryear{{Lemen} et~al.}{{Lemen}
  et~al.}{2012}]{Lemen12}
{Lemen}, J.~R., et~al. 2012, \solphys, 275, 17

\bibitem[\protect\citeauthoryear{{Long} et~al.}{{Long} et~al.}{2017}]{Long17}
{Long}, D.~M., et~al. 2017, \solphys, 292, 7

\bibitem[\protect\citeauthoryear{{Nitta} et~al.}{{Nitta}
  et~al.}{2013}]{Nitta13}
{Nitta}, N.~V., {Schrijver}, C.~J., {Title}, A.~M.,  \& {Liu}, W. 2013, \apj,
  776, 58

\bibitem[\protect\citeauthoryear{{Ofman} \& {Thompson}}{{Ofman} \&
  {Thompson}}{2002}]{Ofman02}
{Ofman}, L.,  \& {Thompson}, B.~J. 2002, \apj, 574, 440

\bibitem[\protect\citeauthoryear{{Parenti}}{{Parenti}}{2014}]{Parenti14}
{Parenti}, S. 2014, Living Reviews in Solar Physics, 11, 1

\bibitem[\protect\citeauthoryear{{Pesnell}, {Thompson}, \&
  {Chamberlin}}{{Pesnell} et~al.}{2012}]{Pesnell12}
{Pesnell}, W.~D., {Thompson}, B.~J.,  \& {Chamberlin}, P.~C. 2012, \solphys,
  275, 3

\bibitem[\protect\citeauthoryear{{Schou} et~al.}{{Schou}
  et~al.}{2012}]{Schou12}
{Schou}, J., et~al. 2012, \solphys, 275, 229

\bibitem[\protect\citeauthoryear{{Shen} et~al.}{{Shen} et~al.}{2014}]{Shen14}
{Shen}, Y., {Liu}, Y.~D., {Chen}, P.~F.,  \& {Ichimoto}, K. 2014, \apj, 795,
  130

\bibitem[\protect\citeauthoryear{{Srivastava} et~al.}{{Srivastava}
  et~al.}{2016}]{Srivastava16}
{Srivastava}, A.~K., {Singh}, T., {Ofman}, L.,  \& {Dwivedi}, B.~N. 2016,
  \mnras, 463, 1409

\bibitem[\protect\citeauthoryear{{Thompson} et~al.}{{Thompson}
  et~al.}{1999}]{Thompson99}
{Thompson}, B.~J., et~al. 1999, \apjl, 517, L151

\bibitem[\protect\citeauthoryear{{Uchida}}{{Uchida}}{1968}]{Uchida68}
{Uchida}, Y. 1968, \solphys, 4, 30

\bibitem[\protect\citeauthoryear{{Vr{\v{s}}nak} \& {Cliver}}{{Vr{\v{s}}nak} \&
  {Cliver}}{2008}]{Vrs08}
{Vr{\v{s}}nak}, B.,  \& {Cliver}, E.~W. 2008, \solphys, 253, 215

\bibitem[\protect\citeauthoryear{{Vr{\v{s}}nak} \& {Luli{\'c}}}{{Vr{\v{s}}nak}
  \& {Luli{\'c}}}{2000}]{Vrs00}
{Vr{\v{s}}nak}, B.,  \& {Luli{\'c}}, S. 2000, \solphys, 196, 157

\bibitem[\protect\citeauthoryear{{Warmuth}}{{Warmuth}}{2011}]{War11}
{Warmuth}, A. 2011, Plasma Physics and Controlled Fusion, 53, 124023

\bibitem[\protect\citeauthoryear{{Warmuth}}{{Warmuth}}{2015}]{Warmuth15}
{Warmuth}, A. 2015, Living Reviews in Solar Physics, 12

\bibitem[\protect\citeauthoryear{{Warmuth} et~al.}{{Warmuth}
  et~al.}{2004}]{War04}
{Warmuth}, A., {Vr{\v{s}}nak}, B., {Magdaleni{\'c}}, J., {Hanslmeier}, A.,  \&
  {Otruba}, W. 2004, \aap, 418, 1117

\bibitem[\protect\citeauthoryear{{White}, {Balasubramaniam}, \&
  {Cliver}}{{White} et~al.}{2013}]{White13}
{White}, S.~M., {Balasubramaniam}, K.,  \& {Cliver}, E. 2013, Technical report
  of Air Force Research Laboratory, 22, 1

\bibitem[\protect\citeauthoryear{{Wills-Davey}, {DeForest}, \&
  {Stenflo}}{{Wills-Davey} et~al.}{2007}]{Wills07}
{Wills-Davey}, M.~J., {DeForest}, C.~E.,  \& {Stenflo}, J.~O. 2007, \apj, 664,
  556

\bibitem[\protect\citeauthoryear{{Zhang} et~al.}{{Zhang}
  et~al.}{2011}]{Zhang11}
{Zhang}, Y., {Kitai}, R., {Narukage}, N., {Matsumoto}, T., {Ueno}, S.,
  {Shibata}, K.,  \& {Wang}, J. 2011, \pasj, 63, 685

\bibitem[\protect\citeauthoryear{{Zheng} et~al.}{{Zheng}
  et~al.}{2020}]{Zheng20}
{Zheng}, R., {Chen}, Y., {Wang}, B.,  \& {Song}, H. 2020, \apj, 894, 139

\bibitem[\protect\citeauthoryear{{Zhukov} \& {Auch{\`e}re}}{{Zhukov} \&
  {Auch{\`e}re}}{2004}]{Zhukov04}
{Zhukov}, A.~N.,  \& {Auch{\`e}re}, F. 2004, \aap, 427, 705

\bibitem[\protect\citeauthoryear{{Zhukov}, {Rodriguez}, \& {de
  Patoul}}{{Zhukov} et~al.}{2009}]{Zhukov09}
{Zhukov}, A.~N., {Rodriguez}, L.,  \& {de Patoul}, J. 2009, \solphys, 259, 73

\bibitem[\protect\citeauthoryear{{Zong} \& {Dai}}{{Zong} \&
  {Dai}}{2017}]{Zong17}
{Zong}, W.,  \& {Dai}, Y. 2017, \apjl, 834, L15

\bibitem[\protect\citeauthoryear{{Zuccarello} et~al.}{{Zuccarello}
  et~al.}{2017}]{Zuca17}
{Zuccarello}, F.~P., {Chandra}, R., {Schmieder}, B., {Aulanier}, G.,  \&
  {Joshi}, R. 2017, \aap, 601, A26

\end{thebibliography}


\end{document}